\pdfoutput=1

\documentclass[prl,superscriptaddress,amsmath,amssymb,twocolumn,nofootinbib,floatfix]{revtex4-2}

\usepackage{graphicx,bm,amsmath}
\usepackage[usenames,dvipsnames]{xcolor}
\usepackage[colorlinks,bookmarks=false,citecolor=NavyBlue,linkcolor=Red,urlcolor=blue,%backref
]{hyperref}

\usepackage[bbgreekl]{mathbbol}
\usepackage{xcolor}
\usepackage[normalem]{ulem}

\def\doi{http://dx.doi.org/}

\newcommand{\be}{\begin{equation}}
\newcommand{\ee}{\end{equation}}
\newcommand{\bea}{\begin{eqnarray}}
\newcommand{\eea}{\end{eqnarray}}

\newcommand{\titleinfo}{Growth of entanglement entropy under local projective measurements}

\begin{document}

%%%%%%%%%%%%%%%%%%%%%%%%%%%%%%%%%%%%%%%%%%%%%%%%%%%%%%%%
\title{\titleinfo}
%%%%%%%%%%%%%%%%%%%%%%%%%%%%%%%%%%%%%%%%%%%%%%%%%%%%%%%%
\author{Michele Coppola}
\altaffiliation{These authors contributed equally to this paper.}
\affiliation{Université de Lorraine, CNRS, LPCT, F-54000 Nancy, France}
\author{Emanuele Tirrito}
\altaffiliation{These authors contributed equally to this paper.}
\affiliation{SISSA, Via Bonomea 265, 34136 Trieste, Italy}
\author{Dragi Karevski}
\affiliation{Université de Lorraine, CNRS, LPCT, F-54000 Nancy, France}
\author{Mario Collura}
\affiliation{SISSA, Via Bonomea 265, 34136 Trieste, Italy}
\affiliation{INFN, Via Bonomea 265, 34136 Trieste, Italy}

\begin{abstract}
Non-equilibrium dynamics of many-body quantum systems 
under the effect of measurement protocols is attracting an increasing amount of attention.
It has been recently revealed that measurements may induce an abrupt change in the scaling-law of the bipartite entanglement entropy, thus suggesting the existence of different non-equilibrium regimes.
However, our understanding of how these regimes appear and whether they survive in the thermodynamic limit is much less established.

Here we investigate these questions on a one-dimensional quadratic fermionic model: this allows us to reach system sizes 
relevant in the thermodynamic sense.
We show that local projective measurements induce a qualitative modification of the time-growth of the entanglement entropy which changes from linear to logarithmic. 
However, in the stationary regime,
the logarithmic behavior of the entanglement entropy does not survive in the thermodynamic limit and, for any finite value of the measurement rate, we numerically show the existence of a single area-law phase for the entanglement entropy.
Finally, exploiting the quasi-particle picture, we further support our results by analysing the fluctuations of the stationary entanglement entropy and its scaling behavior.

%Recently, several studies of quantum many-body time evolution have theoretically proved that frequent measurements induce a transition in a quantum many-body system, which is characterized by a change in the scaling law of the entanglement entropy.
%Here, we investigate, this, the effect of local projective measurements on quantum quench dynamics for a one-dimensional fermionic model. 
%We show that the measurements induce a time logarithmic growth of the entanglement entropy.  
%Moreover, we also study the stationary properties of the entanglement entropy, and, we map out a sharp transition in terms of the measurement rate in spatial space and in time domain, which demonstrates a transition from an area-law with {\color{Red} finite correlation length} to a volume law phase. {\color{Red} Thanks to the quasi- particle picture, we also find the scaling for the fluctuations of the average entanglement entropy. } 

\end{abstract}

\maketitle

\paragraph{Introduction. ---}
Recently, there has been great interest in studying the quench dynamics of isolated quantum many-body systems, where a global parameter of the Hamiltonian is suddenly changed and the initial state is left to evolve unitarily. 
In this scenario the quantum entanglement represents an invaluable tool to access the intrinsic nature of underlying states and their non-equilibrium properties \cite{amico2008entanglement,eisert2010colloquium,laflorencie2016quantum}. 
In the case of unitary evolution, for local short-ranged Hamiltonians,
the spreading of correlation typically scales in time, the front being bounded by a maximum propagation velocity of the information (as predicted by the Lieb-Robinson bound \cite{lieb1972finite}). 
As a consequence, the bipartite entanglement of a semi-infinite subsystem will grow unbounded: for integrable models, where particle excitations are stable and propagate ballistically, the entanglement growth is linear in time, as predicted by the celebrated Cardy-Calabrese quasi-particle picture \cite{calabrese2005evolution,alba2017entanglement,alba2018entanglement,alba2018entanglement2}. 
In this case, the system thermalizes (in a generalised Gibbs sense) and it is characterized by highly entangled eigenstates, i.e. states following an extensive (with the volume) scaling of their entanglement entropy \cite{rigol2008thermalization, nandkishore2015many, abanin2019colloquium}.

Many factors may affect the non-equilibrium dynamics, and the scaling behavior of entanglement entropy could vary in out-of-equilibrium driving
\cite{calabrese2007quantum,von2018operator,rakovszky2019sub,alba2019quantum}.
A paradigmatic example is that of many-body localization (MBL), in which the entanglement transition is driven by the strength of a local disordered potential \cite{basko2006problem,vznidarivc2008many,bardarson2012unbounded,
iyer2013many,kim2013ballistic,huse2014phenomenology,bauer2013area}. 
As a result of avoiding thermalization in the MBL, the stationary state exhibits area-law entropy for the short-entangled systems, and the entanglement entropy grows logarithmically in time, which is in contrast with linear growth in thermalized case \cite{bauer2013area,kjall2014many}.

Recently an alternative way to realize non-thermalizing states has been
proposed by the use of projective measurements that influence the 
entanglement dramatically \cite{cao2018entanglement,skinner2019measurement}. 
In particular, it has been established that quantum systems subjected to both measurements and unitary dynamics offer another class of dynamical behavior described in terms of quantum trajectories \cite{wiseman1996quantum}, and well explored in the context of quantum circuits \cite{li2018quantum,li2019measurement, li2020conformal,
szyniszewski2020universality,zhang2020nonuniversal,zabalo2020critical,shtanko2020classical, jian2020criticality, 
nahum2017quantum,amoschan2019,szyniszewski2019entanglement, chan2019unitary, lavasani2020topological,lavasani2021measurement,block2021measurement,sang2021measurement,shi2020entanglement,lunt2020measurement,sierant2021universal}, quantum spin systems \cite{dhar2016measurement,turkeshi2020measurement,lang2020entanglement,
rossini2020measurement,turkeshi2021measurement,botzung2021engineered, boorman2021diagnosing,fuji2020measurement,ippoliti2021postselection,turkeshi2021measurement2}, trapped atoms \cite{elliott2015multipartite}, and trapped ions \cite{czischek2021simulating,noel2021observation,sierant2021dissipative} .
In this context, the most celebrated phenomenon is the quantum Zeno effect \cite{degasperis1974does,misra1977zeno,peres1980zeno,snizhko2020quantum,biella2021many} according to which continuous projective measurements can freeze the dynamics of the system completely.
%signaling the transition of the entanglement from a volume-law to an area-law. 
This question has been addressed in many-body open systems \cite{cao2018entanglement,alberton2020trajectory,muller2021measurement,goto2020measurement, buchhold2021effective, minato2021fate,PhysRevLett.126.120603} 
whose dynamics is described by a Lindblad master equation
\cite{carollo2019unraveling,vznidarivc2014large,carollo2017fluctuating}.   

In light of these developments, here we study the competition between the unitary dynamics and the random projective measurements in a non-interacting spin-less fermion system.
In particular, we investigate how the bipartite entanglement entropy and its fluctuations are affected by the monitoring of local degrees of freedom in a true Hamiltonian extended model. 

As a main result, we find that the volume-law phase is absent for any measurement rate to sub-extensive entanglement content. In particular, during the initial time-dependent transient, any finite measurement rate induces an abrupt change of the entanglement, whose linear ramp suddenly changes to logarithmic growth. Moreover, we have numerical evidence that the average of the stationary entanglement entropy shows a single transition from the volume- to the area-law phase for any measurement rate in the thermodynamic limit. 
However, for any finite sub-subsystem size, a remnant of a logarithmic scaling is observed, and a characteristic scaling-law at a size-dependent measurement-rate is established.
\vspace{0.2cm}

\paragraph{Protocol. ---}

%Specifically, we have in mind 
Let us consider a quantum many-body 
system in one dimension, whose total Hilbert space $\mathcal{H} = \bigotimes_{j} \mathcal{H}_{j}$, is the tensor product of the single-particle Hilbert spaces $\mathcal{H}_{j}$. 
The system is originally isolated from the environment and
the dynamics obeys the Schr\"odinger equation
$
|\Psi(t)\rangle = \exp\{-i t \hat H\} |\Psi(0)\rangle,
$
where, in our protocol, the initial state is not an eigenstate of the Hamiltonian $\hat H$, and it is typically a very short-correlated state, 
e.g. a product state $|\Psi(0)\rangle = \bigotimes_j |\phi_j\rangle$.

In our protocol, the unitary dynamics is perturbed by random interactions 
with local measuring apparatus: namely, each single local (in real space) 
Hilbert space $\mathcal{H}_{j}$ is coupled for a very short period of time with the environment, and a local observable $\hat O_{j} = \sum_{k=1}^{K} o_{k} \hat P^{(k)}_{j}$ is measured.
Here $o_{k}$ is a possible outcome of the measurements,
and $ \hat P^{(k)}_{j}$ is the projector to the corresponding subspace, with
$\sum_{k=1}^{K} \hat P^{(k)}_{j} = \hat 1_j$.
Given a time step $dt$ and a characteristic rate $ 1/\tau$, 
each single local degree of freedom is independently monitored;
the state $|\Psi\rangle$ is projected according to the Born rule
\be\label{eq:POVM_pure}
|\Psi\rangle \to \frac{\hat P^{(k)}_j |\Psi\rangle}{\sqrt{p_k}}\text{,}
\ee
with probability $p_k = \langle \Psi |  \hat P^{(k)}_{j}|\Psi\rangle $.

In practice, a random number $p \in (0,1]$ is extracted, and a projection to the $k$-th  subspace is performed whether $\sum_{l=1}^{k-1}p_l<p\leq\sum_{l=1}^{k}p_l$.

Under this dynamical protocol, the many-body state $|\Psi(t)\rangle$,
is therefore conditioned by the set of measurement events 
and subsequent outcomes but it remains pure all along the protocol.

Let $\hat{\rho}_i$ denote the density operator for the particular quantum trajectory $\mathcal{T}_i$; $\hat{\rho}_i$ being a projector. Let $\mathcal{O}[\hat{\rho}]$ be a general functional of the density operator. In the following, $\overline{{\mathcal{O}}}$ will denote the average over all the trajectories. In general,
\begin{equation}
    \overline{{\mathcal{O}}^j}=\frac{1}{N}\sum_{i=1}^{N}(\mathcal{O}[\hat{\rho}_i])^j\text{,}\hspace{1cm}\forall j\geq 1
\end{equation}
where $N$ is the number of quantum trajectories.
Let $\overline{\hat{\rho}}=\frac{1}{N}\sum_{i=1}^{N}\hat{\rho}_i$ be the average density operator: $\overline{{\mathcal{O}}}=\mathcal{O}[\overline{\hat{\rho}}]$ only if $\mathcal{O}$ is a linear functional of $\hat\rho$.

Let us mention that, although the stochastic nature of the measurement events remains, the probabilistic outcome of a quantum projective measure can be circumvented by introducing the statistical mixture; indeed, if a measurement is performed but the result of that measurement is unknown, the state is not pure anymore and transforms according to
$
\hat \rho \to \sum_{k=1}^{K} \hat P^{(k)}_j \hat \rho\hat P^{(k)}_j
$,
where at the beginning $\hat \rho(0) = |\Psi(0)\rangle\langle \Psi(0)|$.
The two approaches are indistinguishable as far as we are considering observables which are linear functionals of the density operator $\hat\rho(t)$.

\paragraph{The hopping fermions. ---}
Specifically, we apply our protocol to non-interacting spin-less fermions hopping on a ring with $L$ lattice sites. 
The Hamiltonian with periodic boundary conditions (PBC) reads
\be\label{eq:H_fermions}
\hat H =-\frac{1}{2}\sum_{j=0}^{L-1} \left ( \hat c^{\dag}_{j}\hat c_{j+1} + \hat c^{\dag}_{j+1}\hat c_{j} \right) 
= \sum_{k=-L/2}^{L/2-1} \epsilon_k \, \hat \eta^{\dag}_{k}\hat \eta_{k},
\ee
which is diagonal in terms of the fermionic Fourier modes 
$
\hat \eta_{k} = \frac{1}{\sqrt{L}}\sum_{j=0}^{L-1} e^{-{\rm i} 2\pi k j/L} \hat c{_j}, 
$
with single particle energies $\epsilon_{k} = -\cos(2\pi k /L)$. The Hamiltonian commutes with the total number of particles $\hat N = \sum_{j} \hat n_{j} = \sum_{k} \hat \eta^{\dag}_{k}\hat \eta_{k}$. Due to its quadratic nature, the unitary dynamics preserves the gaussianity of the state, i.e. Wick theorem applies. In practice, in case of closed quantum systems, whenever no measurement occurs, the two-point function $\mathbb{C}_{ij}(t) = \langle \hat c^{\dag}_{i}(t) \hat c_{j}(t) \rangle$ evolves according to $\mathbb{C}(t+s) =\mathbb{R}^{\dag}(s)\hspace{0.1cm}\mathbb{C}(t)\hspace{0.1cm} \mathbb{R}(s)$, where the elements of the matrix $\mathbb{R}(s)$ are
\bea
\mathbb{R}_{mn}(s) 
&=& \frac{1}{L} \sum_{j = -L/2}^{L/2-1} e^{-{\rm i} 2\pi (m-n) j/L - {\rm i} \epsilon_j s} \\
&\sim& {\rm i}^{m-n}J_{m-n}(s), \quad {\rm for} \quad L\sim\infty,
\eea
$J_{k}(z)$ being the Bessel function of the first kind. 

{%Projective measurements could in principle destroy the gaussian property of the state; however, here we only consider operators which are {\it quadratic} in the fermions, and whose measurements do not spoil such a property. }

We focus on a dynamical protocol where
we measure the local occupation number $\hat n_{j} = \hat c^{\dag}_{j} \hat c_{j}$, which is {\it quadratic} in the fermions. Projective measurements could in principle destroy the Gaussian property of the state; however, we shall demonstrate that these particular measurements do not spoil such a property. By hypothesis, $\rho\propto e^{\sum_{ij}\mathbb{M}_{ij}\hat c^{\dag}_i\hat c_j}$ where $\mathbb{M}$ is given coefficient matrix. Due to the spectral decomposition, $\hat n_{j} = 1\cdot\hat P^{(1)}_{j}+0\cdot \hat P^{(0)}_{j}$. Moreover, $\hat n_{j}$ is an hyper-maximal Hermitian operator and thus $\hat 1_j=\hat P^{(1)}_{j}+\hat P^{(0)}_{j}$. In this way, we have just proved that each local number operator is itself a projector ($\hat n_{j} = \hat P^{(1)}_{j}$ and $\hat 1_j-\hat n_{j} = \hat P^{(0)}_{j}$). Secondly, $\hat P^{(1)}_{j}$ and $\hat P^{(0)}_{j}$ can be written as the limit of Gaussian operators, namely $\hat n_j=\lim_{\alpha\to\infty}e^{\alpha \hat n_j}/(e^\alpha-1)$ and $\hat 1_j-\hat n_{j}=\lim_{\alpha\to\infty}e^{-\alpha \hat n_j}$. Finally, $e^{\pm\alpha \hat n_j}e^{\sum_{ij}\mathbb{M}_{ij}\hat c^{\dag}_i\hat c_j}e^{\pm\alpha \hat n_j}=e^{\sum_{ij}\mathbb{K}^{\pm}_{ij}\hat c^{\dag}_i\hat c_j}$ where $\mathbb{K}^{\pm}$ is a new matrix whose elements are given by the Baker-Campbell-Hausdorff formula. Therefore our protocol preserves the Gaussianity of the state.  

Since occupation operators acting on different lattice sites commute, we can apply the following projecting procedure in any arbitrary order; specifically, if at time $t$ the $k$-th site has been measured,
following the prescription in Eq.(\ref{eq:POVM_pure}), if the outcome is $1$ then the state projects as $|\Psi(t)\rangle \to \hat n_{k} |\Psi(t) \rangle/\sqrt{\langle \Psi(t)|  \hat n_{k}|\Psi(t)\rangle}$ otherwise (outcome $0$) the state projects $ |\Psi(t)\rangle \to  (1- \hat n_{k}) |\Psi(t) \rangle/\sqrt{\langle \Psi(t) |1-  \hat n_{k}|\Psi(t)\rangle}$.
The resulting state remaining Gaussian, we can thus focus on the two-point function $\mathbb{C}_{ij}(t)$ which completely characterises the entire system.
The recipe is the following: 
for each time step $dt$ and each chain site $k$, we extract a random number $q_k \in (0,1]$ and only if $q_k\leq dt/\tau$ we take the measurement of the occupation number $\hat n_k$. In such case, we extract another random number $p_k \in (0,1]$: if $p_k \leq \mathbb{C}_{kk}(t) = \langle \hat n_k(t)\rangle$, then thanks to the Wick theorem, the two-point function transforms as
\be
\mathbb{C}_{ij}(t) 
\to 
\delta_{ik}\delta_{jk} + \mathbb{C}_{ij} (t)
- \frac{\mathbb{C}_{ik}(t)\mathbb{C}_{kj}(t)}{\mathbb{C}_{kk}(t)}\text{,}
\ee
otherwise, if $p_k> \mathbb{C}_{kk}(t)$, one obtains
\be
\mathbb{C}_{ij}(t) 
\to 
-\delta_{ik}\delta_{jk} + \mathbb{C}_{ij}(t)
+ \frac{(\delta_{ik}-\mathbb{C}_{ik}(t))(\delta_{jk}-\mathbb{C}_{kj}(t))}{1-\mathbb{C}_{kk}(t)}\text{.}
\ee
Let us mention that, if we lose the result of the measurements,
thus introducing a statistical mixture at every measurement, 
this will definitively spoil the Gaussian nature of the dynamics. 
Therefore, we would lose the great advantage of working with 
a non-interacting theory. For such reason, we will always consider pure-state evolution along quantum trajectories.

%%%%%%%%%%%%%%%%%%%%
\begin{figure}[t!]
\begin{center}
\includegraphics[width=0.5\textwidth]{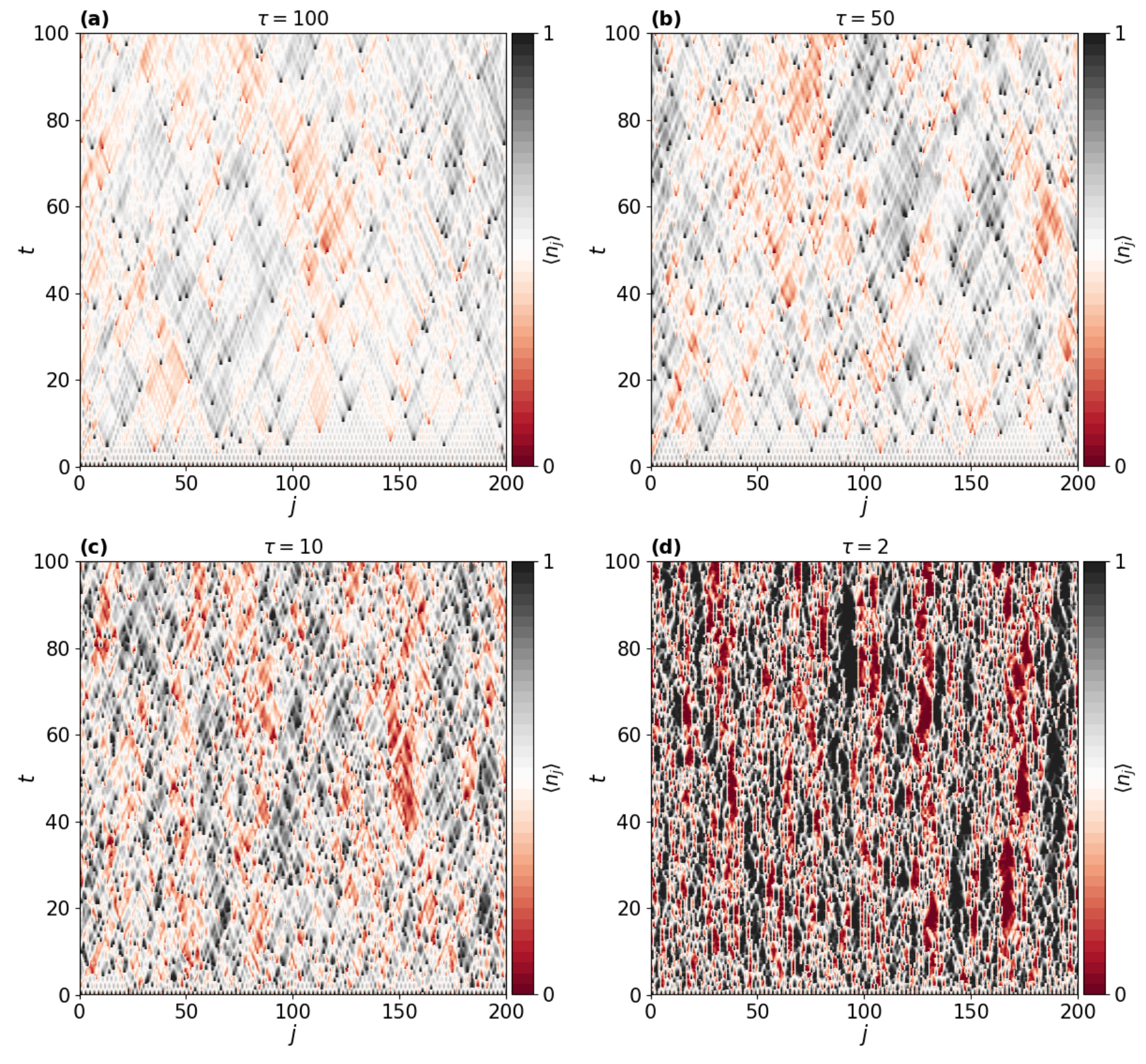}
\caption{\label{fig:nj_vs_tau} \textbf{Evolution of particle density:} The local particle density after quenching the N\'eel state. 
The different panels represent a typical trajectory where random projective measurements of the local occupation $\hat n_j$ occur with different rates $1/\tau$.
}
\end{center}
\end{figure}
%%%%%%%%%%%%%%%%%%%%

In Figure \ref{fig:nj_vs_tau} we show the typical evolution of the 
particle density when starting from the N\'eel product state
$
\prod_{j=0}^{L/2-1} \hat c^{\dag}_{2j} |0\rangle
$
for a system with $L=200$ lattice sites.
Without measurements, the evolution follows the ordinary melting dynamics,
and the states relax (in a local sense) toward the infinite temperature density matrix. Typically, local measurements, when very dilute in time ($\tau \gg 1$),
generate spikes on top of the infinite temperature landscape, 
provided that correlation functions are characterised by a typical finite relaxation time. However, such local excitations, namely  $\hat n_{j}$ or $1-\hat n_{j}$ with almost equal probability, propagates, and survive for ``infinite'' time; indeed, when a local measurement occurs  in the infinite temperature background, the local density at the measured site will relax as $\langle \hat n_{j}(t)\rangle \simeq [1 \pm  J_0(2t)]/2$, with $J_0(2t)\sim t^{-1/2}$; 
moreover, the connected correlation function 
$\langle \hat n_{j}(t)\hat n_{0}(t)\rangle_c = \langle \hat n_{j}(t)\hat n_{0}(t)\rangle - \langle \hat n_{j}(t)\rangle\langle\hat n_{0}(t)\rangle \simeq J^2_j(2t)/4$
spreads ballistically and the front of the light-cone (at $j = 2 t$) 
behaves as $ \langle \hat n_{2t}(t)\hat n_{0}(t)\rangle_c \sim t^{-2/3}$.
Since free quasi-particles have an infinite lifetime, 
they do modify the infinite-temperature landscape at arbitrary distances; 
as a consequence, we may expect that local projective measurements 
should affect the unitary dynamics even for infinitesimally small rate $1/\tau$, 
due to the power-law decay of such ballistically-propagating excitations.

%%%%%%%%%%%%%%%%%%%%%%%%%%%%%%%
\begin{figure}[t!]
\begin{center}
\includegraphics[width=0.5\textwidth]{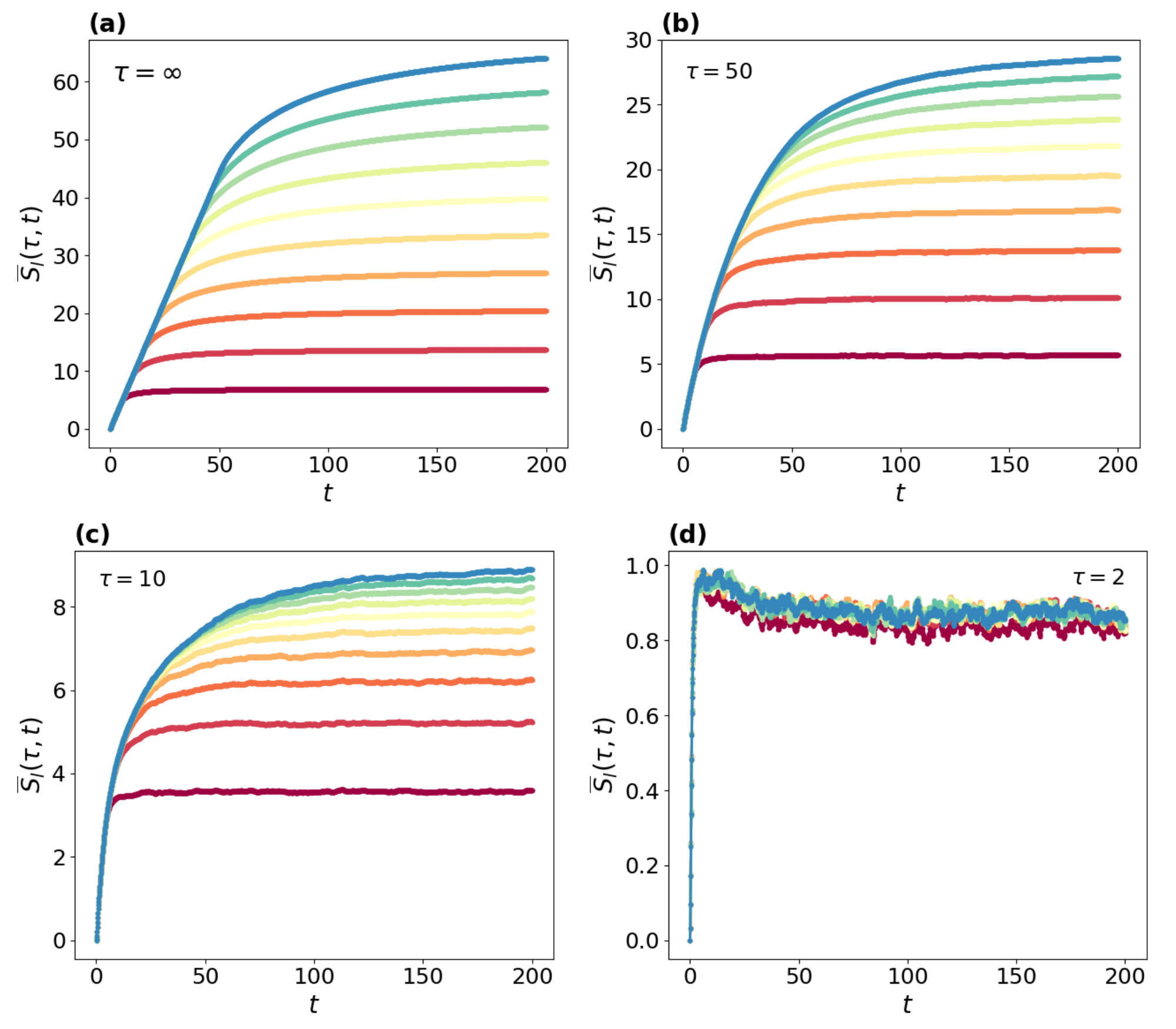}
\caption{\label{fig:S_finite_ell} \textbf{Entanglement Entropy vs time:}
EE after quenching the N\'eel state for a system with size $L=500$. Lines, from bottom to top, represent increasing subsystem sizes  $l \in \{10n : n\in\mathbb{N}\wedge n\leq 10\}$. The results for finite values of $\tau \in \{2,10,50, \infty\}$ have been obtained by averaging over $1000$ different quantum trajectories.}
\end{center}
\end{figure}
%%%%%%%%%%%%%%%%%%%%%%%%%%%%%%
\vspace{0.2cm}
\paragraph{Entanglement entropy dynamics. ---}
One quantity which is definitively affected by the random projective measurements is the bipartite entanglement entropy (EE). 
For a pure state $|\Psi\rangle$, the EE between a subsystem $\mathcal{S}$, and the rest of the system $\mathcal{S^{\star}}$, is given by $S = -{\rm Tr}_{\mathcal{S}} [ \hat \rho_\mathcal{S} \ln \hat \rho_\mathcal{S} ]$,
where $\hat \rho_\mathcal{S} = {\rm Tr}_{ \mathcal{S^{\star}}} |\Psi \rangle \langle \Psi |$ is the reduced density matrix.

In the hopping fermion case, where the dynamical protocol preserves the gaussianity of the state, the time-dependent entropy, for a subsystem consisting of $l$ contiguous lattice sites can be evaluated as
\cite{calabrese2005evolution,vidal2003entanglement,fagotti2008evolution,alba2009entanglement}
\be
S_{l}(t) = 
-\sum_{k} [ \lambda_k(t) \ln \lambda_k(t)
+ (1- \lambda_k(t)) \ln (1 - \lambda_k(t))]\text{,}
\ee
where $\lambda_k(t)$ are the eigenvalues of the subsystem two-point correlation function $\mathbb C(t)|_l$. $\mathbb C(t)|_l$ is an $l\times l$ matrix such that $\mathbb C_{ij}(t)|_l=\mathbb C_{ij}(t)\hspace{0.2cm}\forall i,j\in[0,l)$.

When no measurements occur, the dynamics when starting from the Néel state is typically characterized by a {\it linear increase} for $t \leq l/2$ (quasi-particle velocity $c=1$), followed by a regime where the entropy is saturating toward an {\it extensive} stationary value equal to $l \ln(2)$.
In the opposite case, namely when $\tau \to 0$ and we keep measuring the system everywhere at every time, the state remains completely factorized and the EE is identically vanishing.

In general, a finite rate of random projective measurements should lower the 
entanglement production. However, it is much less clear how this in practice takes place: in particular, are both regimes affected in the same way? Is there any abrupt change in the qualitative behaviour of the entanglement, or this change smoothly depends on the measurements rate $\tau^{-1}$?

We systematically study these questions by analysing the dynamics of the
bipartite EE for different subsystems of sizes $l$, embedded in a system of size $L$.
We performed averages over $200 \div 1000$ different quantum trajectories depending on the specific protocol and system size. At the time $t=0$ the system is prepared in the N\'eel state.

%%%%%%%%%%%%%%%%%%%%%%%%%%%%%%%
\begin{figure}[t!]
\begin{center}
\includegraphics[width=0.45\textwidth]{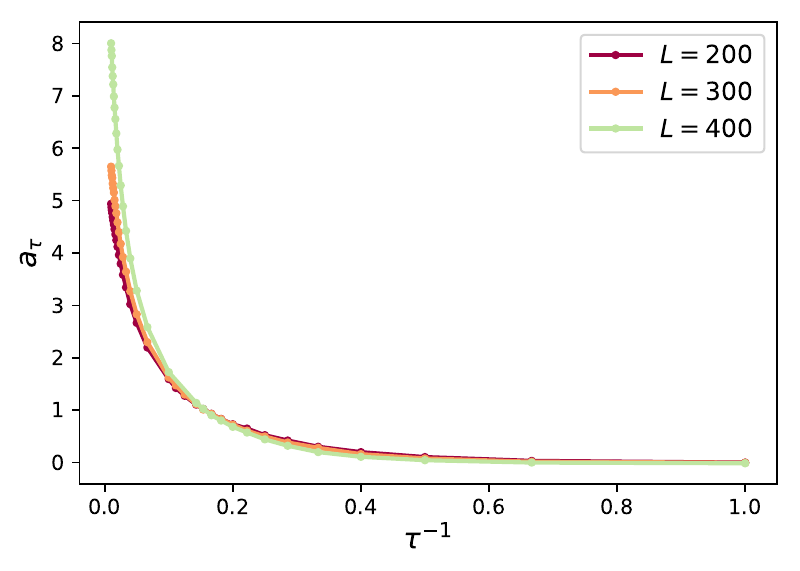}
\caption{\label{fig:S_time_L} 
\textbf{Logarithmic growth: }Logarithmic slope of the EE
extracted from $\overline{S}_l(t) = a_\tau \ln t +b_\tau$ at $l=L/4$ as a function of $\tau$ for different system sizes ($L=200,300,400$). It shows that when $\tau$ is getting larger the expected linear increase of the entanglement is restored. On the contrary, when $\tau\to0$, $a_\tau$ is getting closer to zero suggesting an area-law phase. 
%We consider $1000$ trajectories for any point. 
}
\end{center}
\end{figure}
%%%%%%%%%%%%%%%%%%%%%%%%%%%%%%

In Figure \ref{fig:S_finite_ell}, we show the typical behaviour of the bi-partite EE
for $l \in \{10n : n\in\mathbb{N}\wedge n\leq 10\}$ and system size $L=500$. Maximum time and subsystem sizes have been chosen in such a way that data are not affected by finite-$L$ effects.

For $\tau=\infty$ (Figure \ref{fig:S_finite_ell} (a)), the entropy increases
linearly in time and then saturates at asymptotic values which increase linearly with the subsystem size, thus manifesting the expected extensive behaviour of the stationary EE in accordance with a volume law $l \ln(2)$. 
Decreasing $\tau$, the linear growth of the EE suddenly changes to a logarithmic growth (see Figure \ref{fig:S_finite_ell} (b)-(c)) in accordance to  $ \overline{S}_l(t) = a_\tau \ln t +b_\tau$ which eventually saturates at large time. Finally for very small value of $\tau$ (see Figure \ref{fig:S_finite_ell} (d)), 
we have numerical evidence that the EE shows a rapid saturation to a plateau which is independent of the subsystem size.
Moreover, from the bipartite EE at $l=L/4$ we extract the parameter $a_{\tau}$ by fitting the data with $t\in[0,L/8]$.
In Figure \ref{fig:S_time_L} we show $a_{\tau}$ as a function of the measurement rate $1/\tau$, for different system sizes $L\in\{200,300,400\}$.
As expected, $a_{\tau}$ is growing when $\tau$ is getting larger, eventually diverging for $\tau\to\infty$, so as to restore the expected linear increase of the entanglement when no measurement occurs. On the contrary, for $\tau\to0$, $a_{\tau}$ is vanishing.

Of course, the larger $\tau$, the larger the times and the larger the subsystems have to be in order to appreciate the deviation from the standard linear growth. Notice that, since for smaller $\tau$ much more measures occur, in principle, one needs to take averages over a larger number of quantum trajectories in order to smooth down the random fluctuations.

Interestingly, a finite rate of projective measurements also affects
the scaling of the stationary value of the EE. From a qualitative inspection of the data, the stationary value of the EE undergoes a qualitative change as well: from being extensive when $\tau = \infty$, it shows an area-law scaling for high rates $1/\tau$. However, it is less clear if the area-law behaviour also applies for any finite measurement rate or not. Does it exist another phase between those two asymptotic cases? In other terms, does it exist a critical measurement rate at which we observe a new {\it logarithmic} phase? If yes, can we determine its value? In the following, we go deep to give a definitive answer. 
\vspace{0.2cm}
\paragraph{Stationary entanglement entropy. ---}
We start our analysis of the stationary behavior of the bipartite EE as a function of the subsystem sizes and measurement rates $1/\tau$. Within the zero entanglement when $\tau = 0$ and the volume-law scaling when $\tau=\infty$, we want to study if the stationary EE shows the intermediate logarithmic behaviour when $\tau$ is tuned and the possible existence of a finite critical parameter $\tau_c>0$ which may separate the logarithmic regime from the area-law regime.

To address this question, we inspect the stationary EE as a function of the subsystem size $l$, and different system sizes $L$.
By convention and in order to reduce the fluctuations, we also take the time average over the time window 
$[t_{min},t_{max}]$ wherein the entanglement is almost constant;
where $t_{min} \geq l/2$ and $t_{max} \leq (L-l)/2$
have been chosen so that the entropy is weakly affected by finite-size effects; in fact, in that interval, the EE has essentially entered the stationary regime and is not affected by the motion of particles under PBC on a finite ring. In the following, $\langle \cdot\rangle$ will denote the time average of any functional in that time interval. 
% For this reason, $\langle \overline{S}_l(\tau)\rangle$ represents the stationary value of the average EE for the parameter $\tau$ and the subsystem size $l$. (see Figure \ref{fig:S_vs_l} (a)). 

%%%%% ENTROPY LOG SCALE  vs  ell %%%%%%
\begin{figure}[t!]
\begin{center}
\includegraphics[width=0.45\textwidth]{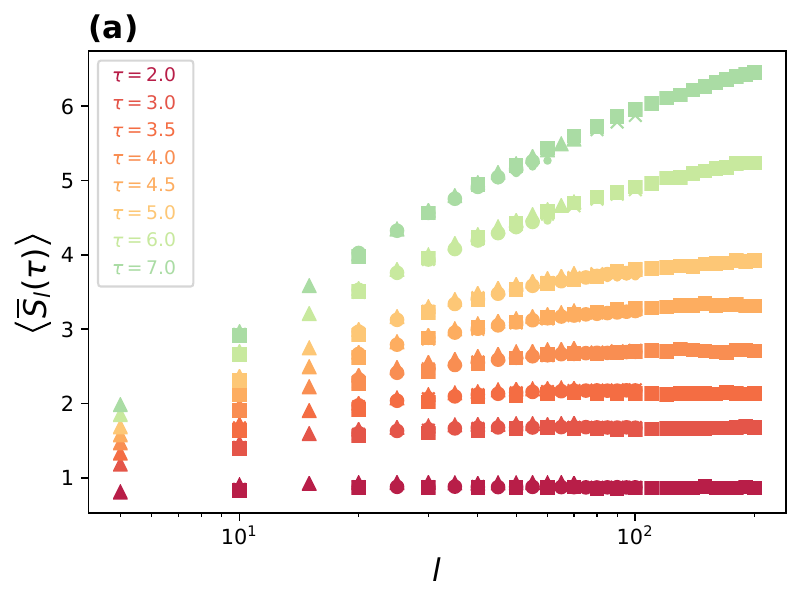}\\
\includegraphics[width=0.45\textwidth]{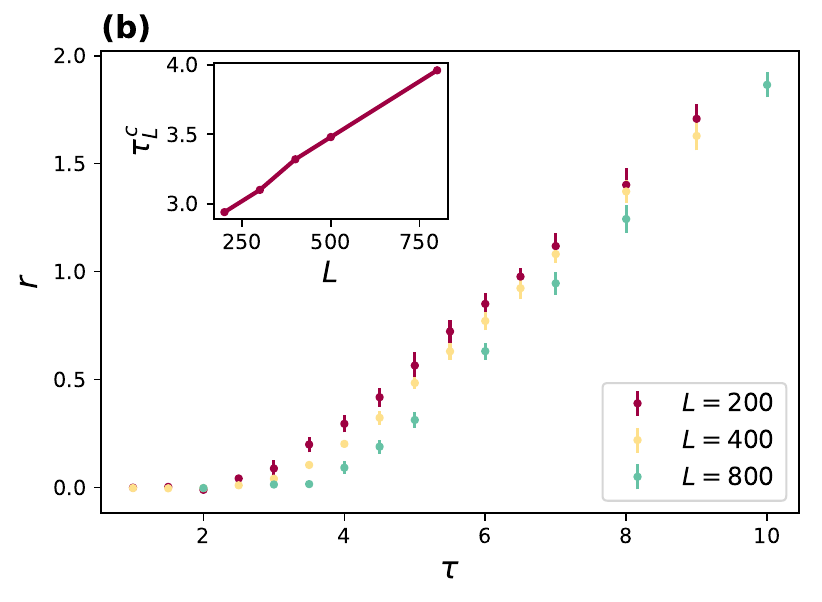}
\caption{\label{fig:S_vs_l} 
\textbf{The stationary entanglement:} 
(a) Stationary EE for $L\in\{200,300,400,500,800\}$ (different symbols) and different measurement rates $1/\tau$ (different colors), as a function of the subsystem size $l\in [1,200]$, is plotted in log-linear scale. 
(b) Logarithmic slope of the stationary EE extracted from different system sizes $L$ as a function of the measurement rates $\tau$. 
The inset shows a finite-size scaling of the critical value $\tau^c_L$ revealing its divergence increasing the system size $L$; see main text for details.
}
\end{center}
\end{figure}
%%%%%%%%%%%%%%%%%%%%%%%%%%%%%%

\begin{figure}[t!]
\begin{center}
\includegraphics[width=0.5\textwidth]{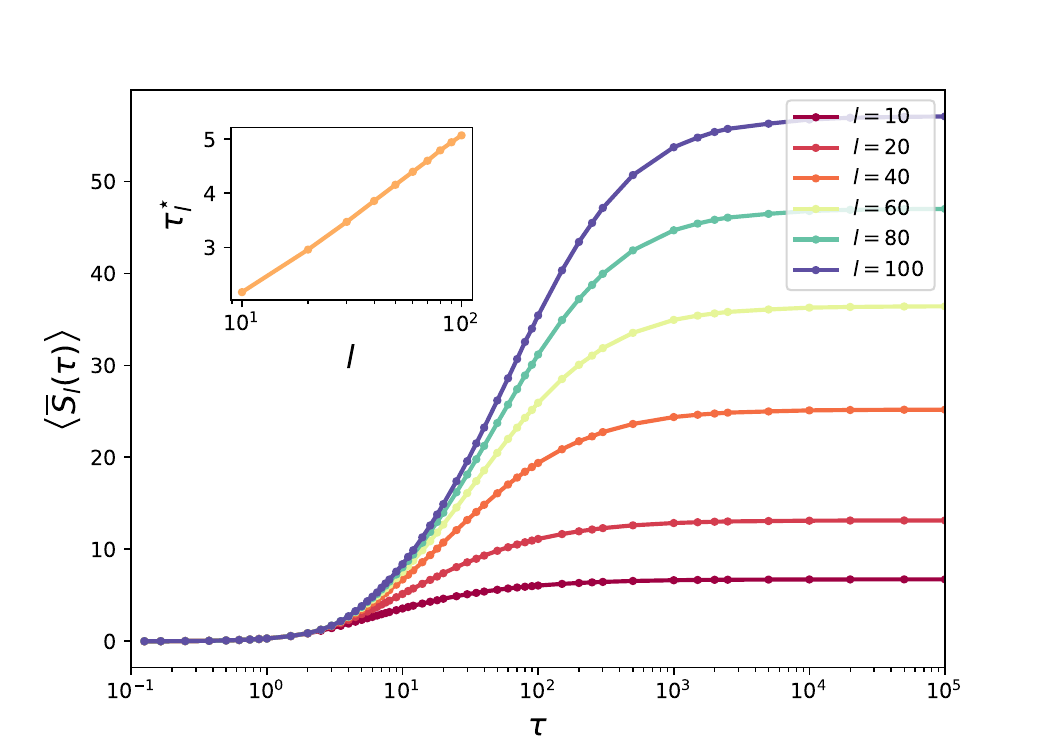}
\caption{\label{fig:inflection_point} 
\textbf{Stationary entanglement entropy vs $\tau$:} 
The stationary EE as a function of $\tau$ for different subsystem sizes $l$ and $L=400$. The entropy is approximately independent on $l$ for small values of the parameter $\tau$. The inset shows the divergence of the inflection point $\tau^{\star}_l$ for $L=400$ as a function of the subsystem size $l$.} 
\end{center}
\end{figure}

By an analysis of the data, we conclude that the entanglement entropy saturates to a constant value independent on $l$ for a sufficiently small value of $\tau$; in other words, the measurement rate is so high that the time-evolved system cannot escape from a short correlated state. We can safely say that the system is in a Zeno-like regime in which the measurements have suppressed the entanglement, giving rise to an area-law scaling. 

In order to verify if the asymptotic scaling acquires a logarithmic dependence with the subsystem size for increasing values of the parameter $\tau$, we make use of a linear fit $r \ln(l) + k$ of the data with $l\in[l_{min},l_{max}]$ and we extract the parameter $r$: it gives an estimate of the asymptotic logarithmic slope of the entanglement, namely $l \partial_l \langle\bar S_{l}(\infty)\rangle$, as a function of the measurement rates $1/\tau$. Here, $l_{min}$ and $l_{max}$ have been chosen depending on the system size $L$ in order to stay in the correct regime.

In Figure \ref{fig:S_vs_l} (b) we plot the best fit parameter $r$ for $\tau \in [1,10]$, and different system sizes $L\in\{200,400,800\}$. This quantity is an indicator of a possible sharp transition between different regimes in the asymptotic scaling of the EE.
Similarly to what has been observed for the scaling of the entropy in the time-dependent regime, the logarithmic slope $r$ for every system size $L$ decreases when going toward $\tau = 0$. In particular, for every size $L$, we can identify a critical value $\tau^c_L$ such that, for $\tau<\tau^c_L$, $r$ shows a fast convergence toward zero. To estimate $\tau^c_L$ we perform a best fit of $r$ for every system size whose intercept with the axis $r=0$ gives the critical value $\tau^c_L$ separating the area-law phase from the logarithmic regime. Indeed, if the limit $\lim_{L\to\infty} \tau^c_L= \tau^c $ converges to a finite value, the {\it logarithmic phase} manifests also in the stationary regime and $\tau_c$ captures the phase transition point between logarithmic and area-law phase. 
However, the data in the inset of Figure \ref{fig:S_vs_l} (b) suggest that $\tau^c_L$ is linearly growing with $L$, and therefore the only stationary-EE phase that survives in the thermodynamic limit is indeed the area-law phase.

This statement is strongly supported by the study of the EE as a function of $\tau$ for different subsystem sizes as shown in Figure \ref{fig:inflection_point}. 
For very high measurement rates $1/\tau$, the stationary EE is independent on the subsystem size $l$ with a very good approximation. As $\tau$ increases, the stationary EE becomes $l$-dependent and changes its concavity at $\tau = \tau^{\star}_l$. In particular, our data show a logarithmic growth of the inflection point $\tau^{\star}_l$ with the subsystem size $l$. This observation allows us to introduce a correlation length $\xi(\tau)$ which increases exponentially with $\tau$ and affects very much the behaviour of the stationary EE. In fact, if $\xi(\tau)\ll l$ then only the chain's sites close to the boundary of the subsystem are correlated with the rest of the quantum system. For this reason, the EE is $l$-independent. As $\xi(\tau)$ gets larger and larger, more and more sites are involved in generating correlation with the rest of the chain. 
When $l \sim \xi(\tau)$ the EE shows a logarithmic scaling with the subsystem size; however this region moves in the parameter space $\tau-l$ such that it eventually tends to infinity in the thermodynamic limit, thus disappearing. 
Finally, for $\xi(\tau)\gg l$ the entire subsystem contributes to the EE 
and this essentially results in a volume-law behaviour.

\begin{figure}[t!]
\begin{center}
\includegraphics[width=0.5\textwidth]{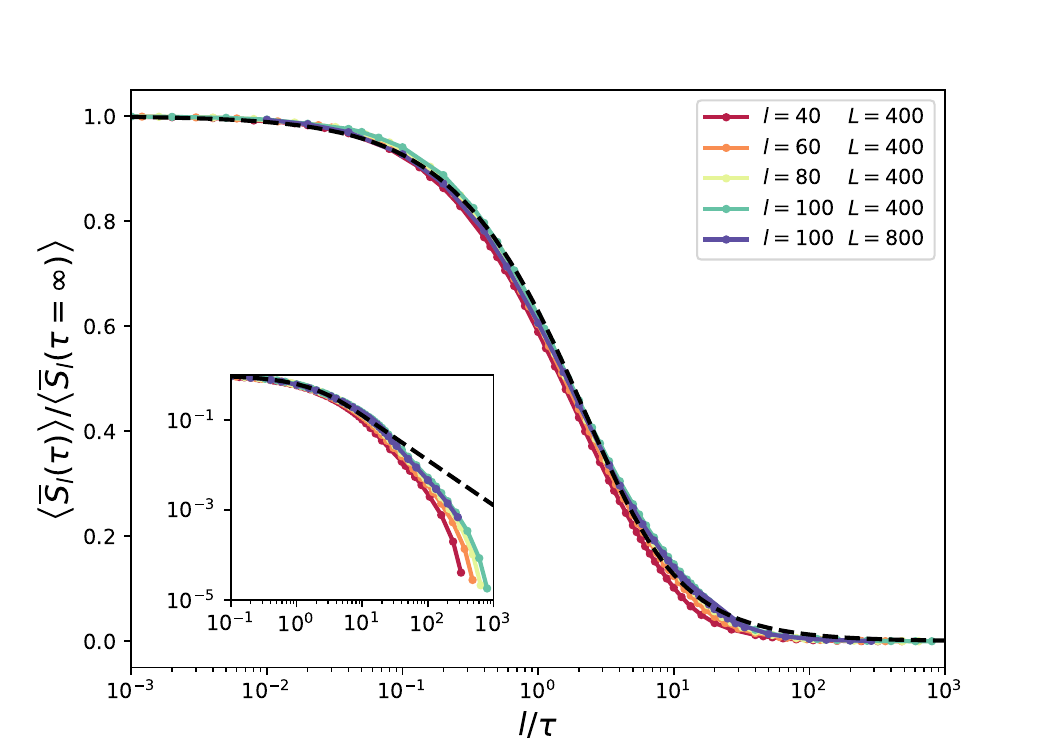}
\caption{\label{fig:scal} 
\textbf{Scaling of the stationary entanglement entropy:} 
Scaling of the stationary EE: $\lambda=l/\tau$ on the x-axis; $\langle\overline{S}_l(\tau)\rangle/\langle\overline{S}_l(\tau=\infty)\rangle$ on the y-axis. The dashed line is the function $f(\lambda)$ given by the GHD. Note that $\langle\overline{S}_l(\tau)\rangle$ has been computed by taking the average in $[t_{min},t_{max}]$: it means that $\langle\overline{S}_l(\tau=\infty)\rangle$ is very well approximated by $l \ln(2)$ only if $l\ll L$. The inset shows the same data in log-log scale to emphasize the differences between the models for high rates. }  
\end{center}
\end{figure}
\vspace{0.2cm}
\paragraph{Entanglement entropy scaling. ---}
As well known, the Generalized Hydrodynamic (GHD)~\cite{PhysRevLett.117.207201, castro2016emergent,bulchandani2017solvable, 10.21468/SciPostPhysLectNotes.18, alba2021generalizedhydrodynamic} makes use of a quasi-particle picture \cite{alba2018entanglement,alba2017entanglement, alba2018entanglement2} to explain qualitatively the behaviour of the entanglement dynamics. 
Let $x_1$ and $x_2$ be two general points of the chain: they define the subsystem $\mathcal{I}=[x_1,x_2]$ of interest for the EE ($|x_1-x_2|=l<L$). 
For weakly-entangled and excited quantum states, the EE under unitary time evolution is \cite{cao2018entanglement}
\begin{equation}\label{free}
    S_l(t)=\int_{-\pi}^{\pi}\frac{dk}{2\pi}\int_{\mathcal{Q}_{k,t}}dx\hspace{0.1cm}s(x-v(k)t,k,0)\text{,}
\end{equation}
where $v(k)=\sin(k)$ is the group velocity of the quasi-particles, $s(x,k,t)$ represents the contribution to the EE for a pair of quasi-particles at positions $x$ and $x-2v(k)t$, and $\mathcal{Q}_{k,t}=\{x\in \mathcal{I}\hspace{0.2cm}|\hspace{0.2cm}x-2v(k)t \not\in \mathcal{I}\}$.

It is interesting to notice that, in the continuous limit ($dt\to 0$), the time evolution of the average density operator is given by the Lindblad equation (where the jump operators coincide with the local number operators) of which the Stochastic Schr\"odinger Equation (SSE) is an unravelling. In other terms, 
\begin{equation}\label{Lindblad}
    \frac{d\overline{\hat{\rho}}^{\mbox{\tiny CL}}}{dt}=-{\rm i}[\hat{H},\overline{\hat{\rho}}^{\mbox{\tiny CL}}]-\frac{1}{\tau}\sum_{j=1}^{L}\big[\hat{n}_j,[\hat{n}_j,\overline{\hat{\rho}}^{\mbox{\tiny CL}}]\big]\text{,}
\end{equation}
where $\overline{\hat{\rho}}^{\mbox{\tiny CL}}$ is the average density operator in the continuous limit (CL). 

In the CL, we can use the quasi-particle picture and the postulates for the entanglement growth in presence of continuous measurements, where $1/\tau$ represents the monitoring rate of each quasi-particle.\footnote{Note that the rate used in \cite{cao2018entanglement} is two times bigger than the rate we deduce from \eqref{Lindblad}.} As put forward in Ref. \cite{cao2018entanglement}, the idea is based on the possibility that the ballistic motion may be stopped by a random measure event which destroys a pair of quasi-particles and generates a new excitation which starts spreading from the position of one of the two old partners with the same probability $1/2$. The new quasi-particles travel with random momenta $\pm k$ where $|k|$ is chosen uniformly in $[0,\pi]$. 
Let $\overline{S}_l^{\mbox{\tiny CL}}(\tau,t)$ denote the average EE in the CL. Using the prescriptions in \cite{cao2018entanglement} with the physical measurement rate $1/\tau$, we get 
\begin{equation}\label{av}
    \overline{S}_l^{\mbox{\tiny CL}}(\tau,t\to\infty)=\ln(2)\int_0^\infty \frac{dy}{\tau}\hspace{0.1cm} e^{-y/\tau}\int^{\pi}_{-\pi}\frac{dk}{2\pi}\int_{\mathcal{Q}_{k,y}}dx,
\end{equation}
whose asymptotic behavior reads
\begin{equation}\label{f1}
    \frac{\overline{S}_l^{\mbox{\tiny CL}}(\tau,t\to\infty)}{S_{l}(\tau=\infty)}=f(\lambda)\text{,}\hspace{0.48cm}       f(\lambda)\sim\begin{cases}1\hspace{0.7cm}\lambda\ll 1\\
   \lambda^{-1}\hspace{0.3cm}\lambda\gg 1
   \end{cases}
\end{equation}
where $\lambda=l/\tau$, and $S_{l}(\tau = \infty)=\ln(2) l$ is the asymptotic value of the EE under free time evolution. 
Since our protocol differs from the one in Ref.\cite{cao2018entanglement},
it is worth investigating whether our recipe agrees with their scaling result.
Actually, assuming $dt\ll 1$, the EE in our discrete model is very well captured by the CL description when $dt/\tau\ll1$. Even if we expect to have a good prediction by the GHD only for $\tau\gg1$, we see in Figure \ref{fig:scal} that the agreement is excellent in a much wider range of measurement rates. Figure \ref{fig:scal} also shows the same ratio for different chain sizes $L\in\{400,800\}$ to emphasize that it is weakly affected by finite-size effects. However, using bigger chains ensures better agreement between data and theoretical predictions, as expected.

\begin{figure}[t!]
\begin{center}
\includegraphics[width=0.5\textwidth]{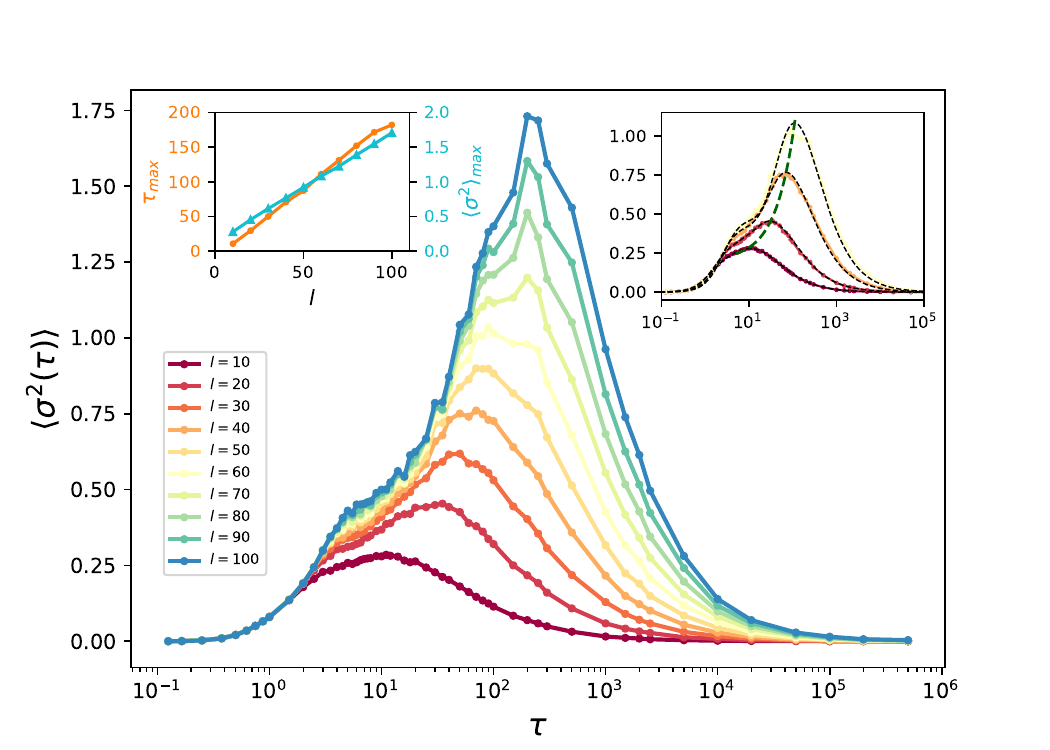}
\caption{\label{fig:vcal} 
\textbf{Fluctuations of the stationary entanglement entropy:} 
Fluctuations of the stationary EE as a function of the parameter $\tau$ for $L=400$. The absolute maximum point $\langle\sigma^2\rangle_{max}$ and its position $\tau_{max}$ increase linearly with the subsystem size $l$, as shown in the inset on the left. The inset on the right shows some fit functions (dashed lines) with their data; the green curve interpolates their maximum points. } 
\end{center}
\end{figure}
\vspace{0.2cm}

\paragraph{Stationary entanglement entropy fluctuations  ---} 
In order to further support the results of the previous sections, 
we decided to analyse the EE fluctuations $\langle\sigma^2_l(\tau)\rangle=\langle\overline{S^2}_l(\tau)\rangle-\langle\overline{S}_l(\tau)^2\rangle$.
From the numerical results, we see that: {\it(i)} the variance is $l$-independent 
for very high measurement rates; {\it(ii)} it approximately decays as 
$\sim 1/\tau$ for very low rates (see Figure \ref{fig:vcal}). 
The behaviour at low $\tau$ is not surprising: for very high rates, we are close to the Zeno regime and then we do expect that also higher momenta of the EE are size independent. 
The behaviour at high values of $\tau$ may be easily understood 
if we look at the proprieties of the Poisson distribution,
as detailed below. Indeed, suppose $\tau$ is large enough in order to satisfy $Ldt/\tau\ll 1$; 
in this case, the probability to have multiple measurement events after each time step $dt$ is approximately zero. Under this assumption, the measurement process becomes a Poisson process and $\kappa=TL/\tau$ is the average number of measurements in the time interval $[0,T]$. 
It follows that
\begin{equation}
\overline{S^n}_l(\tau,T)=\sum_j\mathcal{P}(j)\mathcal{A}_l(j,n,T),
\end{equation}
where $\mathcal{P}(j)=\kappa^j e^{-\kappa}/j!$ is the probability to take $j$ measurements in $[0,T]$ and $\mathcal{A}_l(j,n,T)$ is the weighted average of the $n$-th momentum over all the quantum trajectories which can be generated in $[0,T]$ with fixed number of measurements $j$. The variance is thus given by
\begin{equation}
\begin{aligned}
    \sigma_l^2(\tau,T)&=\overline{S^2}_l(\tau,T)-(\overline{S}_l(\tau,T))^2\\
    &\sim\mathcal{F}_l(T)\kappa(\tau,T)\text{,}\hspace{1cm}\kappa\ll 1
\end{aligned}
\end{equation}
where $\mathcal{F}_l(T)=\mathcal{A}_l(1,2,T)-2\mathcal{A}_l(0,1,T)\mathcal{A}_l(1,1,T)+\mathcal{A}_l(0,1,T)^2$. Of course, we are interested in computing the stationary variance and thus $T$ has to be large enough. It is interesting to notice that, in this regime, $\sigma_l^2(\tau,hT)=\sigma_l^2(\tau,T)$ with $h>1$. In fact, we can show that $\mathcal{F}_l(hT)=\mathcal{F}_l(T)/h$ and $\kappa(\tau,hT)=h\kappa(\tau,T)$. 

\begin{figure}[t!]
\begin{center}
\includegraphics[width=0.5\textwidth]{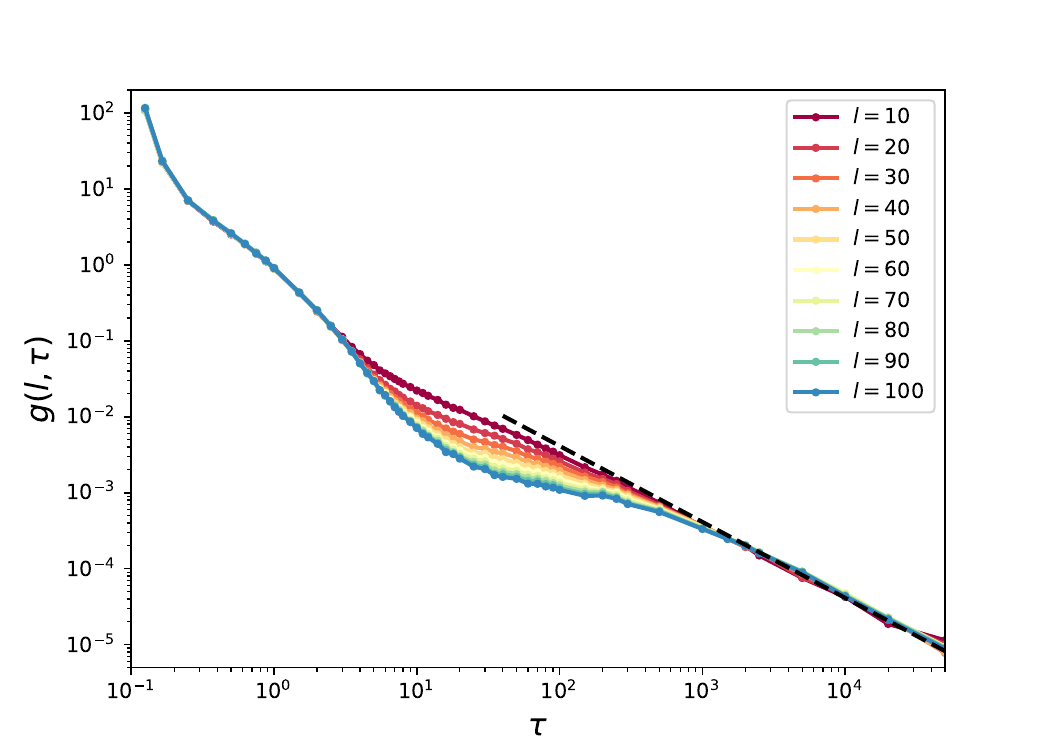}
\caption{\label{fig:vcal2} 
\textbf{Scaling of the fluctuations for small measurement rates:} 
The ratio $g(l,\tau)=\langle\sigma^2_l(\tau)\rangle/\langle\overline{S}_l(\tau)\rangle^2$ as a function of the parameter $\tau$ for $L=400$. The dashed line represents the asymptotic behaviour obtained by theoretical argues; see main text for details.  } 
\end{center}
\end{figure}

The term $\mathcal{A}_l(1,n,T)$ represents the contribution for one single measurement and then it is not surprising that it can be written in terms of small perturbations with respect to the $0$-measurement case. In other terms, $\mathcal{F}_l(T)$ is approximately proportional to $l^2$ and thus the ratio $g(l,\tau) = \langle\sigma^2_l(\tau)\rangle/\langle\overline{S}_l(\tau)\rangle^2$ is essentially $l$-independent and proportional to $1/\tau$ for very low measurement rates. This behaviour is emphasized in Figure \ref{fig:vcal2}. 

In addition to the interesting asymptotic behaviours, the variance shows a double-peak structure (see Figure \ref{fig:vcal}) which might be evidence of the existence of two different processes generating fluctuations. Note that this double-peak structure also affects the ratio $g(l,\tau) = \langle\sigma^2_l(\tau)\rangle/\langle\overline{S}_l(\tau)\rangle^2$, as shown in Figure \ref{fig:vcal2}. 

Our data suggest that the position of the peak on the left scales logarithmically with the subsystem size $l$: this is not surprising because, in that regime, fluctuations reflect the behaviour of the inflection point $\tau^*_l$ of the EE which also scales logarithmically. Despite the fact that the number of simulated trajectories is not sufficient to proceed with a thorough analysis, it seems that the peak's position of the absolute maximum $\tau_{max}$ and its value $\langle\sigma^2\rangle_{max}=\langle\sigma^2(\tau_{max})\rangle$ increase linearly with the subsystem size $l$, as shown in the inset of the Figure \ref{fig:vcal}. In order to estimate the maximum points and their positions, we fit the data with a linear combination of two functions obtained by the square of the relation \eqref{av} and the square of the average contribution to the EE of the new pairs of particles randomly created by measurements (which appears in the argument of the integral \eqref{av}). By optimizing the parameters of these fit functions, we obtain a very good interpolation of the data, as shown in the inset of the Figure \ref{fig:vcal}. This might suggest that it is possible to describe the processes generating fluctuations making use of the quasi-particle picture. 

\vspace{0.2cm}
\paragraph{Discussion and conclusion. ---} 
In this work, we investigated the quantum quench dynamics in a free fermion chain under projective measurements of occupation numbers. By computing the EE of the system during the time evolution, we found that the entanglement shows a logarithmic growth in time before reaching the stationary value. {%we found that when the rate of the measurements decreases the scaling of the entanglement changes from an area-law to a logarithmic phase.} 
Furthermore, thanks to the experimental progress, this logarithmic regime that
emerges for finite sizes system, can be also addressed in laboratory \cite{bloch2008many,islam2015measuring,elben2020mixed,brydges2019probing}.

Moreover, we also investigated the properties of the stationary EE as a function of the measurement rate $1/\tau$ and we studied a volume- to area-law transition that emerges for any value of $\tau$. Finally, we studied the scaling of the stationary EE, the fluctuations, and the ratio between the variance and the square of the stationary EE as a function of the measurement rate, finding out a linear asymptotic behaviour. 
We found a very intriguing phenomenon where the EE fluctuations are generated by two distinct processes
which are both qualitatively captured by the quasi-particle picture.

\bibliographystyle{apsrev4-1}
\bibliography{bibliography}

%\begin{thebibliography}{99}

%\end{thebibliography}
\end{document}